\documentclass[twocolumn,showpacs,preprintnumbers,amsmath,amssymb]{revtex4}
\usepackage{hyperref}
\usepackage{revsymb}
\usepackage{graphicx}% Include figure files
\usepackage{dcolumn}% Align table columns on decimal point
\usepackage{bm}% bold math

\begin{document}
\draft
\title{Nearly maximal violation of the Mermin-Klyshko inequality with
multimode entangled coherent states}
\author{Jian-Qi Sheng, Li-Hua Lin, Zhi-Rong Zhong}
\email{zhirz@fzu.edu.cn}
\author{Shi-Biao Zheng}
\email{t96034@fzu.edu.cn}
\address{Fujian Key Laboratory of Quantum Information and Quantum Optics,
College of Physics and Information Engineering, Fuzhou University, Fuzhou,
Fujian 350108, China}

\begin{abstract}
Entangled coherent states for multiple bosonic modes, also referred to as
multimode cat states, not only are of fundamental interest, but also have
practical applications. The nonclassical correlation among these modes is
well characterized by the violation of the Mermin-Klyshko inequality. We
here study Mermin-Klyshko inequality violations for such multi-mode
entangled states with rotated quantum-number parity operators. Our results show
that the Mermin-Klyshko signal obtained with these operators can approach
the maximal value even when the average quantum number in each mode is only
1, and the inequality violation exponentially increases with the number of
entangled modes. The correlations  among the rotated parities of the entangled 
bosonic modes are in distinct contrast with those among the displaced parities, 
with which a nearly maximal Mermin-Klyshko inequality violation requires the size 
of the cat state to be increased by about 15 times.
\end{abstract}

\vskip 0.5cm
\maketitle
\narrowtext

\section{INTRODUCTION}

Quantum mechanics predicts many counterintuitive phenomena in microscopic
world, among which nonlocality is particularly appealing. According to
quantum mechanics, when the wavefunctions of two objects are entangled,
there exist nonlocal correlations between the features of these two objects,
irrespective of their spatial distance. Einstein, Podolsky, and Rosen (EPR)
rejected this point of view, and thought that the results of measurements
performed on two particles should be mutually independent when their
separation is sufficiently large~[1]. Based on this locality condition, they
concluded that quantum mechanical description of nature is incomplete.
Despite the debate between Einstein and Bohr [2], it was not until Bell's
discovery in 1964 that it was realized that predictions of quantum mechanics
conflict with local realism for some entangled states [3]. In the framework
of local realism, Bell developed an inequality, which sets an upper bound on
the statistical correlations of the results of measurements performed on two
distant spin-1/2 particles. According to quantum mechanics, this bound is
exceeded with proper settings of the measurement orientations when these two
particles are prepared in a highly entangled state before the measurements.

Various forms of Bell inequality have been formulated, among which Clauser,
Horne, Shimony and Holt (CHSH) version [4] is most famous and has been
widely used for experimental test of quantum nonlocality [5-7]. The
Bell-CHSH inequality has been generalized in different directions. On one
hand, an inequality, referred to as the Mermin-Klyshko (MK) inequality, was
derived for systems composed of multiple spin-1/2 particles [8,9], which
shows that the conflict between local hidden theories and quantum mechanics
becomes stronger as the number of entangled particles increases.
Experimental violation of the MK inequality was reported with four-photon
entanglement [10], and was analyzed in the context of ion traps~[11]. On the
other hand, Banaszek and W\"{o}dkiewicz constructed a Bell-CHSH-type
operator for two harmonic oscillators with continuous variables, and
formulated an inequality in phase space [12,13], which was later generalized
to maximize the Bell signal [14,15]; however, with this formalism the
inequality is not maximally violated by the original EPR state. In contrast,
based on the so-called pseudospin operators introduced by Chen et al.~[16],
the original EPR state can maximally violate the inequality, but the
experimental measurements remains challenging. We have constructed a new
Bell operator [17], with which the Bell signal for a two-mode cat state of
moderate size can approach the upper bound of quantum mechanics. The issue
of quantum nonlocality associated with multimode continuous variable states
has also been addressed [18-23]. In particular, Jeong et al. [18] have
discussed the quantum nonlocality associated with entangled coherent states
for three bosonic modes using joint quasiprobability distribution function
in phase space. With this method, there are 12 variables to be optimized to
find the global maximum value of the MK signal, which is a difficult task.
In Ref.~[18], only some local maximum values were numerically found. For the
Greenberger-Horne-Zeilinger-type (GHZ-type) [24] entangled coherent states,
also referred to as 3-mode Schr\"{o}dinger cat states [25-28], the
obtained MK signal increases with the amplitude $\alpha $ of the coherent
state components, reaching a maximal value of 3.6 when $\alpha \rightarrow
\infty $; this maximum is still lower than the quantum-mechanical upper
bound of 4 for a three-partite system by 10\%. With this method, the number
of the variables to be optimized increases with the number of entangled
bosonic modes, which makes it extremely difficult to obtain the optimal MK
signal when more bosonic modes are involved. We note that, multimode
entangled coherent states are intriguing both from the fundamental view
point and for practical applications because the components forming these
highly nonclassical states have classical analogs and can be used for
fault-tolerant quantum computation [29].

In this manuscript, we analyzed the MK inequality violations for GHZ-type
entangled coherent states of $n$ bosonic modes based on effective rotated
parity correlations proposed in Ref. [17]. Our results show that the MK
signal quickly increases with $\alpha $, and approaches the
quantum-mechanical upper bound when $\alpha $ is only about 1. This is in
distinct contrast with the result obtained in Ref. [18], which shows the MK
signal for the 3-mode cat state at $\alpha =1$ is even significantly below
the lower bound for confirming true 3-partite entanglement. We further show
that the MK signal obtained with the joint quasiprobability distribution
function can also approach the quantum-mechanical upper bound for a
sufficiently large cat size. However, with this approach the size of the cat
state needs to be increased by about 15 times to obtain a nearly maximal MK
inequality violation.

The paper is organized as follows. In Sec. 2, we construct the MK operator
for $n$-mode cat states with effective rotated parity operators. With this
formalism, the correlations among the bosonic modes are in almost perfect
analogy to those among $n$ qubits prepared in a GHZ state even for $\alpha
\sim 1$, in contrast with the results obtained in continuous-variable
representation. In Sec. 3, We detailedly analyze the MK inequality violation
for the 3-mode cat state based on these correlations. Analytical results
qualitatively show that the size of the cat state does not need to be large
for the MK signal to approach the quantum-mechanical upper bound. We confirm
this prediction with numerical simulation, which demonstrates that a nearly
maximal inequality violation can be obtained even when $\alpha $ is only
about 1. We find that, when using joint quasiprobability distribution
function, the average quantum number in each mode needs to be increased by
about 15 times to obtain the same degree of violation. In Sec. 4, we
investigate the MK signals for the 4-mode and 5-mode cat states. Numerical
simulations show that, for these cases the MK signals can also nearly
approach the corresponding quantum-mechanical upper bounds even when the
average quantum number in each mode is only 1. Conclusions are presented in
Sec. 5.

\section{CONSTRUCTION OF THE MK OPERATOR FOR N-MODE CAT STATES}

The $n$-mode Schr\"{o}dinger cat state under consideration is defined as the
equal superposition of all the $n$ bosonic modes being in the coherent state
$\left\vert \alpha \right\rangle $ and all in $\left\vert -\alpha
\right\rangle $,

\begin{eqnarray}
\left\vert \psi \right\rangle _{n-cat}&&={\cal N}_{n}(\left\vert \alpha
\right\rangle _{1}\left\vert \alpha \right\rangle _{2}\left\vert \alpha
\right\rangle _{3}...\left\vert \alpha \right\rangle _{n}\nonumber \\&&+\left\vert -\alpha
\right\rangle _{1}\left\vert -\alpha \right\rangle _{2}\left\vert -\alpha
\right\rangle _{3}...\left\vert -\alpha \right\rangle _{n})£¬
\end{eqnarray}
where ${\cal N}_{n}=\left[ 2+2e^{-2n\left\vert \alpha \right\vert ^{2}}%
\right] ^{-1/2}$ and the subscripts $j=1$ to $n$ label these modes. When $%
\left\vert \alpha \right\vert ^{2}\gg 1$, the two coherent states $%
\left\vert \alpha \right\rangle $ and $\left\vert -\alpha \right\rangle $
are approximately orthogonal and can be considered two logic states of a
qubit. With this encoding, the n-mode Schr\"{o}dinger cat state is in
analogy with the GHZ state of $n$ qubits
\begin{equation}
\left\vert \psi \right\rangle _{\mathrm{GHZ}}=\frac{1}{\sqrt{2}}\left( \left\vert
\uparrow \right\rangle _{1}\left\vert \uparrow \right\rangle _{2}\left\vert
\uparrow \right\rangle _{3}...\left\vert \uparrow \right\rangle
_{n}+\left\vert \downarrow \right\rangle _{1}\left\vert \downarrow
\right\rangle _{2}\left\vert \downarrow \right\rangle _{3}...\left\vert
\downarrow \right\rangle _{n}\right) .
\end{equation}
According to quantum mechanics, when $n$ qubits are prepared in a GHZ state,
there are nonlocal correlations among the outcomes of measurements
individually performed on them. The conflict between the predictions of
quantum mechanics and the results allowed by local realism becomes stronger
as the number of entangled qubits increases, as evidenced by the violations
of the MK inequality. The MK operator for a multipartite entangled state is
recursively defined as
\begin{equation}
O_{k}=\frac{1}{\sqrt{2}}\left[ O_{k-1}\left( \sigma _{k,a_{k}}+\sigma
_{k,a_{k}^{^{\prime }}}\right) +O_{k-1}^{^{\prime }}\left( \sigma
_{k,a_{k}}-\sigma _{k,a_{k}^{^{\prime }}}\right) \right] ,
\end{equation}%
starting with $O_{1}=\sigma _{1,a_{1}}$. Here $\sigma _{k,a_{k}}$ and $%
\sigma _{k,a_{k}^{^{\prime }}}$ represent two-valued observables for the $k$%
th party, with $a_{k}$ and $a_{k}^{^{\prime }}$ denoting the corresponding
measurement settings, and $O_{k}^{^{\prime }}$ is obtained from $O_{k}$ by
exchanging all $a_{j}$ and $a_{j}^{^{\prime }}$ ($j\leq k$). For a
multi-qubit system, $a_{k}$ is a unit vector and $\sigma _{k,a_{k}}$
corresponds to the Pauli operator along the direction $a_{k}$. In the
framework of local hidden variable (LHV) theories, one can assign a value of
$1$ or $-1$ to $\sigma _{k,a_{k}}$ and $\sigma _{k,a_{k}^{^{\prime }}}$, so
that $O_{k}=\pm \sqrt{2}O_{k-1}$ or $O_{k}=\pm \sqrt{2}O_{k-1}^{^{\prime }}$%
. As a result, the outcome of each measurement on $O_{n}$ is $2^{(n-1)/2}$
or $-2^{(n-1)/2}$, which implies that the MK signal within a LHV model,
defined as $S_{n}^{l}=\left\vert \left\langle O_{n}\right\rangle \right\vert
_{LHV}$, satisfies $S_{n}^{l}\leq D_{n}=2^{(n-1)/2}$. To show the violation
of the MK inequality by the $n$-qubit GHZ state, we set all of $a_{k}$ and $%
a_{k}^{^{\prime }}$ are on the xy-plane, and for simplicity, use the
notation $\sigma _{k}(\phi _{k})$ to replace $\sigma _{k,a_{k}}$, with $\phi
_{k}$ being the angle between $a_{k}$ and the $x$ axis. Set%
\begin{eqnarray}
\phi _{1} &=&0,\phi _{1}^{^{\prime }}=\pi /2, \\
\phi _{k} &=&-\pi /4,\phi _{k}^{^{\prime }}=\pi /4,\text{ }k\neq 1.
\nonumber
\end{eqnarray}
Then we have
\begin{equation}
O_{n}=2^{n-1}\left\vert \uparrow \right\rangle _{1}\left\langle \downarrow
\right\vert \otimes \left\vert \uparrow \right\rangle _{2}\left\langle
\downarrow \right\vert ...\otimes \left\vert \uparrow \right\rangle
_{n}\left\langle \downarrow \right\vert +\mathrm{H.c}.,
\end{equation}
where $\mathrm{H.c.}$ denotes the Hermitian conjugate. According to quantum
mechanics, the expectation value of $O_{n}$ for the ideal $n$-qubit GHZ
state is $S_{n}=\left\langle O_{n}\right\rangle _{QM}=2^{n-1}$, which
exceeds the bound imposed by the local realism by a factor of $2^{\left(
n-1\right) /2}$, indicating the violation of the MK inequality exponentially
grows with the  number of entangled qubits.

We first analyze the MK inequality violation for the $n$-mode cat state
based on the effective cat state qubit rotation operators. For the $k$th cat
state qubit, the quantum-number parity operator $P_{k}=(-1)^{a_{k}^{\dagger
}a_{k}}$ acts as $\sigma _{k,x}$, where $a_{k}^{\dagger }$ and $a_{k}$ is
the quantum-number rising and lowering operators. We note the other
components of the Pauli operators in the xy-plane for the cat state qubits
can be operationally constructed by combining the parity operator and the
effective rotation operator, defined as [17]
\begin{equation}
R_{k,z}(\phi _{k})=D_{k}^{\dagger }(\alpha )G_{k}(\phi _{k})D_{k}(\alpha ),
\end{equation}
where
\begin{equation}
G_{k}(\phi _{k})=\left\vert 0\right\rangle _{k}\left\langle 0\right\vert
e^{i\phi _{k}}+\sum_{n=1}^{\infty }\left\vert n\right\rangle
_{k}\left\langle n\right\vert
\end{equation}
is the phase gate, and $D_{k}(\alpha )=e^{\alpha a_{k}^{\dagger }-\alpha
^{\ast }a_{k}}$ denotes the displacement operator. When $\left\vert
\left\langle \alpha \right\vert \left. -\alpha \right\rangle \right\vert
^{2}\ll 1$, $R_{k,z}(\phi _{k})\left\vert \alpha \right\rangle \simeq
\left\vert \alpha \right\rangle $ and $R_{k,z}(\phi _{k})\left\vert -\alpha
\right\rangle \simeq e^{i\phi _{k}}\left\vert -\alpha \right\rangle $.
Therefore, $R_{k,z}(\phi _{k})$ is effectively equivalent to a rotation of
the $k$th cat state qubit around the z axis of the Bloch sphere by an angle $%
\phi _{k}$. The observables needed for test of the MK inequality can be
expressed in terms of this kind of rotation operators and the parity
operator
\begin{equation}
\sigma _{k}(\phi _{k})=R_{k,z}(\phi _{k})P_{k}R_{k,z}^{\dagger }(\phi _{k}).
\end{equation}%
The effective rotated parity operator $\sigma _{k}(\phi _{k})$ corresponds
to the Pauli operator along the axis with an angle $\phi _{k}$ to the x axis
of the cat state qubit. For a finite value of $\alpha $, the coherent states
$\left\vert \alpha \right\rangle _{j}$ and $\left\vert -\alpha \right\rangle
_{j}$ are not strictly orthogonal, so that the $n$-mode cat state is not
perfectly equivalent to the $n$-qubit GHZ state. As a consequence, the MK
signal $S_{n}$ is smaller than $2^{(n-1)/2}$. However, the overlap between $%
\left\vert \alpha \right\rangle _{j}$ and $\left\vert -\alpha \right\rangle
_{j}$, $e^{-4\left\vert \alpha \right\vert ^{2}}$, is quite small even for a
moderate value of $\alpha $. For example, for $\alpha =\sqrt{2}$ this
overlap is only about $2.5\times 10^{-4}$. Therefore, the MK signal $S_{n}$
for the $n$-mode cat state of a moderate size, obtained with the effective
cat state qubit rotation operators, can approximate the result for an $n$%
-qubit GHZ state.

The test of quantum nonlocality in phase space is based on the measurement
of displaced quantum-number parity observables, defined as
\begin{equation}
P_{k}(\beta _{k})=D_{k}^{\dagger }(\beta _{k})P_{k}D_{k}(\beta _{k}).
\end{equation}%
In this formalism, $P_{k}(\beta _{k})$ and $P_{k}(\beta _{k}^{^{\prime }})$
corresponds to $\sigma _{k,a_{k}}$ and $\sigma _{k,a_{k}^{^{\prime }}}$, and
the recursive definition of the MK operator can be written as
\begin{equation}
Q_{k}=Q_{k-1}\left[ P_{k}(\beta _{k})+P_{k}(\beta _{k}^{^{\prime }})\right]
+Q_{k-1}^{^{\prime }}\left[ P_{k}(\beta _{k})-P_{k}(\beta _{k}^{^{\prime }})%
\right] .
\end{equation}
The expectation value of each term of $Q_{n}$ is proportional to the joint
Wigner function (quasiprobability distribution function) for these $n$
bosonic modes at the corresponding point in phase space. For a limited value
of $\alpha $, the displacement operation $D_{k}(\beta _{k})$ moves the
corresponding bosonic mode out of the logic space of the corresponding cat
state qubit, and is not equivalent to a qubit rotation. Therefore, the
obtained MK signal is expected to be smaller than that based on the
effective cat qubit rotation operators.

\section{VIOLATIONS OF THE MK INEQUALITY WITH THE 3-MODE CAT STATE}

To quantitatively show the violations of MK inequality based on the
above-mentioned rotated parity correlations, we will derive analytic
expressions for the quantum mechanical expectation values of the MK
observable $O_{n}$ and then perform numerical simulations. We first consider
the 3-mode cat state, for which the MK operator based on the effective
rotated parity correlations among the three modes is given by
\begin{eqnarray}
&&O_{3} =\sigma _{1}(0)\sigma _{2}(-\pi /4)\sigma _{3}(\pi /4)+\sigma
_{1}(0)\sigma _{2}(\pi /4)\sigma _{3}(-\pi /4)\nonumber \\&&+
\sigma _{1}(\pi /2)\sigma _{2}(-\pi /4)\sigma _{3}(-\pi /4)-\sigma
_{1}(\pi /2)\sigma _{2}(\pi /4)\sigma _{3}(\pi /4).\nonumber \\&&
\end{eqnarray}
The MK signal, defined as the absolute value of the expectation value of $
O_{3}$, is%
\begin{eqnarray}
S_{3}&&=\vert E_{3}(0,-\pi /4,\pi /4)+E_{3}(0,\pi /4,-\pi /4)+ \nonumber \\&&E_{3}(\pi
/2,-\pi /4,-\pi /4)-E_{3}(\pi /2,\pi /4,\pi /4)\vert ,
\end{eqnarray}
where%
\begin{eqnarray}
E_{3}(\phi _{1},\phi _{2},\phi _{3}) &=&{\cal N}_{3}^{2}\left\{
e^{-6\left\vert \alpha \right\vert ^{2}}+\prod_{k=1}^{3}K_{\alpha ,\alpha
}(\phi _{k})\right.  \nonumber \\
&&\left. +\left[ \prod_{k=1}^{3}K_{\alpha ,-\alpha }(\phi _{k})+c.c.\right]
\right\} ,
\end{eqnarray}
with
\begin{eqnarray}
K_{\alpha ,\alpha }(\phi _{k}) &=&-e^{-2\left\vert \alpha \right\vert
^{2}}+ 2e^{-2\left\vert \alpha \right\vert ^{2}}\cos \phi
_{k}+\nonumber\\&& 2e^{-4\left\vert \alpha \right\vert ^{2}}(1-\cos \phi _{k}),\nonumber\\
K_{\alpha ,-\alpha }(\phi _{k}) &=&e^{-i\phi _{k}}+e^{-4\left\vert \alpha
\right\vert ^{2}}\left( 1-e^{-i\phi _{k}}\right) .
\end{eqnarray}
When $e^{-2\left\vert \alpha \right\vert ^{2}}\ll 1$, the dominant term of $
E_{3}(\phi _{1},\phi _{2},\phi _{3})$ is $2{\cal N}^{2}\cos (\phi _{1}+\phi
_{2}+\phi _{3})$. The ratio of this term to the corresponding correlation
for a GHZ state is $2{\cal N}_{3}^{2}$, which approximates $1$ even for a
very moderate value of $\alpha $. For example, it is about $0.9975$ for $%
\alpha =1$. This allows a nearly maximal violation of the MK inequality with
the average quantum number in each mode being only about 1.

To confirm the validity of the rotated parity operators for revealing the
quantum nonlocality among the three bosonic modes, we first perform
numerical simulations of the four correlations $E_{3}(0,-\pi /4,\pi /4)$, $
E_{3}(0,\pi /4,-\pi /4)$, $E_{3}(\pi /2,-\pi /4,-\pi /4)$, and $E_{3}(\pi
/2,\pi /4,\pi /4)$ as functions of $\alpha $. As shown in Eqs. (13) and
(14), these correlations are independent of the argument of the complex
amplitude $\alpha $, which is taken to be a positive real number for
simplicity. The results are respectively shown in Fig. 1(a), (b), (c), and
(d), where the solid lines denote the correlations of the 3-mode cat states,
while the dotted lines represent the corresponding correlations for the
equally-weighted classical mixture of the two components $\left\vert \alpha
\right\rangle _{1}\left\vert \alpha \right\rangle _{2}\left\vert \alpha
\right\rangle _{3}$ and $\left\vert -\alpha \right\rangle _{1}\left\vert
-\alpha \right\rangle _{2}\left\vert -\alpha \right\rangle _{3}$, which are
given by
\begin{equation}
E_{3}^{m}(\phi _{1},\phi _{2},\phi _{3})=\frac{1}{2}\left[ e^{-6\left\vert
\alpha \right\vert ^{2}}+\prod_{k=1}^{3}K_{\alpha ,\alpha }(\phi _{k})\right]
.
\end{equation}
For $\alpha =0$, each mode is in the vacuum state $\left\vert 0\right\rangle
$, which is not affected by the rotation operator $R_{k,z}^{\dagger }(\phi
_{k})$. Consequently, the rotated parity operator $\sigma _{k}(\phi _{k})$
is equivalent to the parity operator $P_{k}$, which has a definite value of
1, so that each correlation is 1. In this case, all correlations are
completely classical as there is no entanglement. As $\alpha $ increases,
for both $\left\vert \alpha \right\rangle $ and $\left\vert -\alpha
\right\rangle $ the difference between the occupations between the even and
odd quantum-number states quickly decreases, and consequently the
correlations $E_{3}^{m}(\phi _{1},\phi _{2},\phi _{3})$ for the classical
mixture drop fast, and approach 0 when $e^{-2\left\vert \alpha \right\vert
^{2}}\ll 1$. In contrast, for this case the 3-mode cat state exhibits nearly
perfect correlations, like the three-qubits GHZ state $\left\vert \psi
\right\rangle _{GHZ}$, which is the eigenstate of each of the four terms in $
O_{3}$, with the eigenvalues corresponding to the first three terms being 1
and that associated with the last term being $-1$. This is in distinct
contrast with the two-mode case, where the correlations used to construct
the Bell signal are not perfect even when $\alpha \rightarrow \infty $; the
absolute value of each correlation tends to $\sqrt{2}/2$ in this limit [17].
These reuslts unambiguously demonstrate that, when $e^{-2\left\vert \alpha
\right\vert ^{2}}\ll 1$, the correlations associated with the 3-mode cat
state almost completely arise from the quantum coherence between $\left\vert
\alpha \right\rangle _{1}\left\vert \alpha \right\rangle _{2}\left\vert
\alpha \right\rangle _{3}$ and $\left\vert -\alpha \right\rangle
_{1}\left\vert -\alpha \right\rangle _{2}\left\vert -\alpha \right\rangle
_{3}$, which is responsible for the entanglement among the three modes.

\begin{figure}
\centering
\includegraphics[width=3.5in]{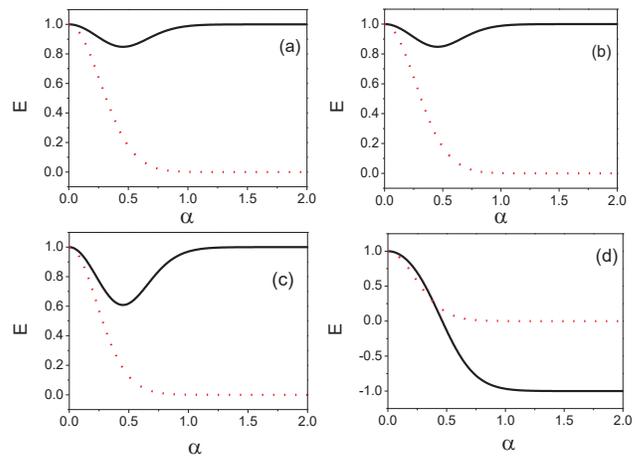}
\vspace{-0.2cm}
\caption{(Color online) Numerical simulations of correlations for the 3-mode
cat state as functions of $\alpha $: (a), $E_{3}(0,-\pi /4,\pi /4)$; (b), $%
E_{3}(0,\pi /4,-\pi /4)$; (c), $E_{3}(\pi /2,-\pi /4,-\pi /4)$; (d), $%
E_{3}(\pi /2,\pi /4,\pi /4)$. The solid and dotted lines corrrespond to the
correlations for the cat state and the equally-weighted classical mixture of
the two components $\left\vert \alpha \right\rangle _{1}\left\vert \alpha
\right\rangle _{2}\left\vert \alpha \right\rangle _{3}$ and $\left\vert
-\alpha \right\rangle _{1}\left\vert -\alpha \right\rangle _{2}\left\vert
-\alpha \right\rangle _{3}$, respectively. }
\end{figure}

The MK signal ($S_{3}$) for the 3-mode cat state as a function of $\alpha $
is shown in Fig. 2, with the blue solid line representing the simulated
result while the black dotted line denoting the bound imposed by local
hidden variable theories. The result shows that $S_{3}$ first drops below 2,
and then quickly increases with $\alpha $. This non-monotonous behavior is
similar to the two-mode case [17]. When $\alpha <0.266$, $E_{3}(\pi /2,-\pi
/4,-\pi /4)$ and $E_{3}(\pi /2,\pi /4,\pi /4)$ almost have the same value
and thus cancel each other out, so that $S_{3}$ is approximately equal to
the sum of the other two correlations, each of which slightly decreases as $%
\alpha $ increases in this regime. Consequently, $S_{3}$ drops as $\alpha $
increases in this regime. When $\alpha >0.266$, the difference between $%
E_{3}(\pi /2,-\pi /4,-\pi /4)$ and $E_{3}(\pi /2,\pi /4,\pi /4)$ increases
quickly, whose contribution makes $S_{3}$ rise quickly. $S_{3}$ reaches the
classical upper bound 2 at $\alpha =0.36$, and exceeds the lower bound $2\sqrt{2
}$ for true 3-partite entanglement [30] when $\alpha >0.581$. When $\alpha $
further increases, it quickly approaches the quantum-mechanical upper bound
4, which implies that a nearly maximal MK inequality violation can be
obtained with a small cat state. For example, when $\alpha =1$, $S_{3}\simeq
3.916$, very close to upper bound 4. This can be interpreted as follows. The
two coherent states $\left\vert \alpha \right\rangle $ and $\left\vert
-\alpha \right\rangle $ of each bosonic mode are approximately orthogonal to
each other and can act as the basis states of a logic qubit even for a
moderate value of $\alpha $. The bosonic mode is well restricted in the
logic space of the cat state qubit under the application of the effective
cat state qubit rotation operator $R_{k,z}(\phi _{k})$. Consequently, the
rotated parity operator $\sigma _{k}(\phi _{k})$ is approximate to the
corresponding Pauli operator, and the 3-mode cat state is approximately an
eigenstate of each of the four correlation operators in Eq. (11).

As noted in Ref. [18], for the MK signal based on the joint Wigner function,
there are 6 complex displacement parameters, $\beta _{k}$ and $\beta
_{k}^{^{\prime }}$ with $k=1$ to $3$, corresponding to 12 variables to be
optimized to obtain the maximal MK signal for each value of $\alpha $, which
is a difficult task. Instead of the global maximum value, a local maximum
value for each value of $\alpha $ was numerically obtained in Ref. [18],
which monotonously increases with $\alpha $, and approaches 3.6, instead of
4, when $\alpha \rightarrow \infty $. We note that the upper bound 4 can be
reached in this limit with the choice $\beta _{1}=0$, $\beta _{1}^{^{\prime
}}=i\pi /8\alpha $, $\beta _{k}=-i\pi /16\alpha $, $\beta _{k}^{^{\prime
}}=i\pi /16\alpha $, with $k>1$. This is due to the fact that the displaced
parity operators are equivalent to the corresponding Pauli operators of the
cat state qubits when $\alpha \rightarrow \infty $, so that the MK
inequality can be maximally violated in this limit. Based on these displaced
parity correlations, the MK signal at $\alpha =3.824$ is equal to the result
of Fig. 2 at $\alpha =1$, which implies that with these correlations the
average quantum number in each mode, $\stackrel{-}{n}=\left\vert \alpha
\right\vert ^{2}$, should be increased by about 15 times for approaching the
maximal violation. According to the result of Ref. [18], the MK\ signal at $%
\alpha =1$ is only about 2.5, which is significantly below the lower bound
for confirming true 3-partite entanglement. This is in stark contrast with
the result obtained with our framework, where the upper bound allowed by
3-partite entanglement can be approached for $\alpha \simeq 1$.

\begin{figure}
\centering
\includegraphics[width=3.5in]{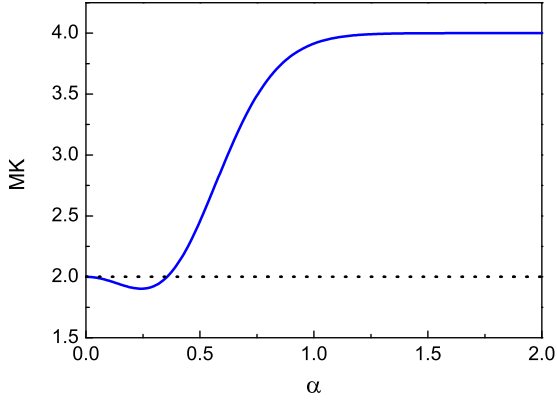}
\vspace{-0.2cm}
\caption{(Color online) MK signal for the 3-mode cat state as a function $%
\alpha $. The solid line represents the numerical result calculated by Eq.
(12), and the dotted line denotes the maximum allowed by local realism. }
\end{figure}

\section{MK INEQUALITY VIOATIONS WITH 4- and 5-MODE CAT STATES}

The MK signal for the 4-mode cat state obtained with the effective rotated
parity operators is given by
\begin{eqnarray}
&& S_{4} =\frac{1}{\sqrt{2}}[-E_{4}(0,-\pi /4,-\pi /4,-\pi /4)+\nonumber \\&& 3E_{4}(0,-\pi
/4,-\pi /4,\pi /4)  +3E_{4}(0,-\pi /4,\pi /4,\pi /4)-\nonumber \\&&E_{4}(0,\pi /4,\pi /4,\pi /4)-  E_{4}(\pi
/2,\pi /4,\pi /4,\pi /4)  -\nonumber \\&& 3E_{4}(\pi /2,\pi /4,\pi /4,-\pi /4)+ 3E_{4}(\pi /2,\pi /4,-\pi /4,-\pi/4) \nonumber \\&&+ E_{4}(\pi /2,-\pi /4,-\pi /4,-\pi /4)].
\end{eqnarray}
where
\begin{eqnarray}
E_{4}(\phi _{1},\phi _{2},\phi _{3},\phi _{4}) &&={\cal N}_{4}^{2}\left\{
e^{-8\left\vert \alpha \right\vert ^{2}}+\prod_{j=1}^{4}K_{\alpha ,\alpha
}(\phi _{j})\right.  \nonumber \\
&&\left. +\left[ \prod_{j=1}^{4}K_{\alpha ,-\alpha }(\phi _{j})+c.c.\right]
\right\} .
\end{eqnarray}
The blue solid line in Fig. 3 represents the rescaled MK signal $R_{4}$, as
a function of the amplitude $\alpha $ of the coherent state components. Here
$R_{4}$ is defined as the ratio of the MK signal to the upper bound ($2\sqrt{
2}$) allowed by local hidden variable models. The results show $R_{4}$ exceeds the classical
upper bound 1 when $\alpha >0.311$, and surpasses the lower bound $2$ for true
4-partite entanglement when $\alpha >0.616$. The maximal value $2\sqrt{2}$ can
be approached for a moderate value of $\alpha $. For example, $R_{4}\simeq
2.758$ when $\alpha =1$. Based on displaced parity operators, the MK signal
reaches this value only when $\alpha $ is above $3.94$.
\begin{figure}
\centering
\includegraphics[width=3.5in]{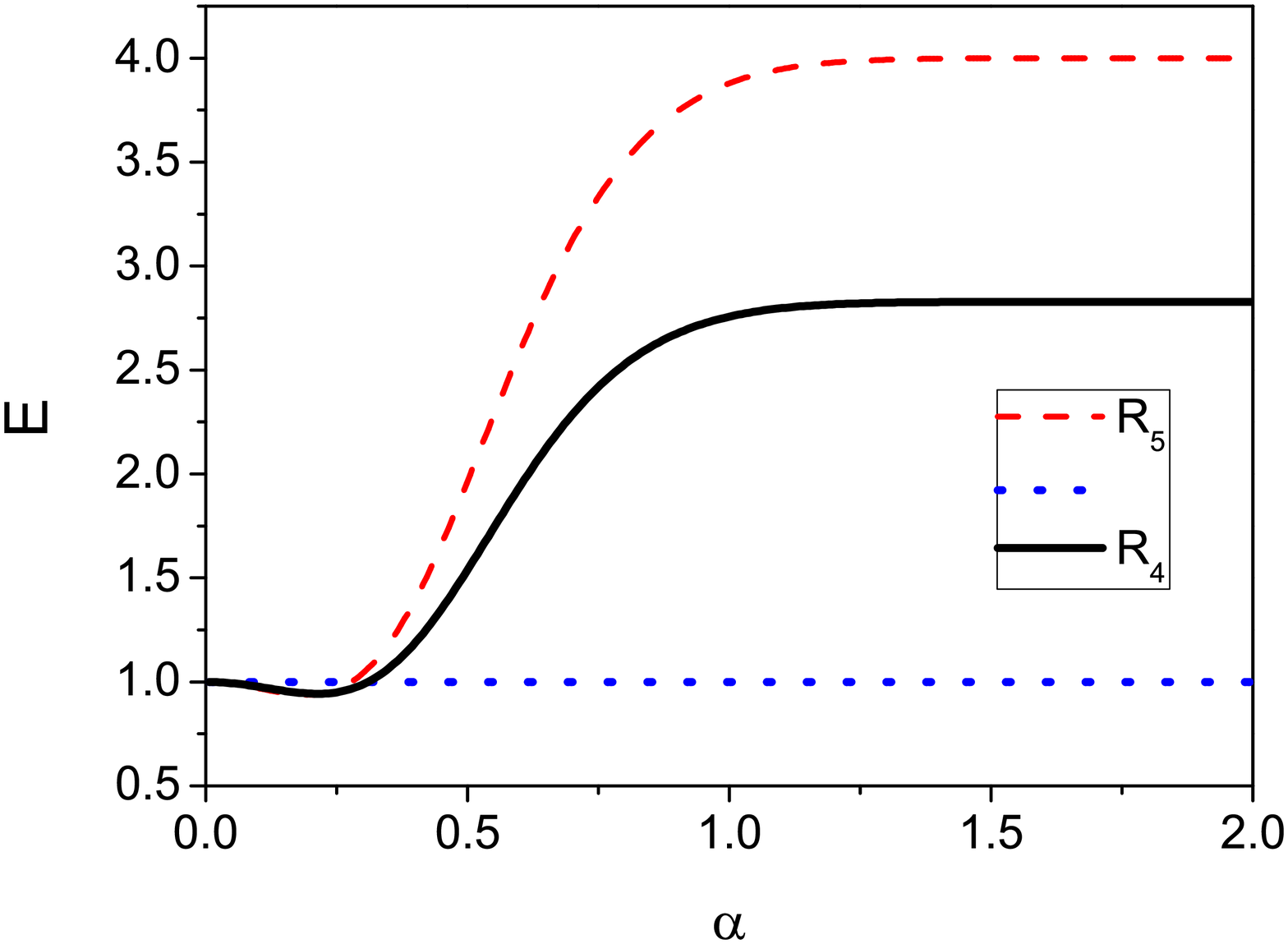}
\vspace{-0.2cm}
\caption{(Color online) Rescaled MK signals for 4- and 5-mode cat states as
functions of $\alpha $. The blue solid line represents the result for the 4-mode
case, the red dashed line corresponds to the 5-mode case, and the black dot
line denotes the maximum allowed by local realism. The rescaled MK signal is
defined as $R_{n}=S_{n}/D_{n}$, where $S_{n}$ is the calculated MK signal
for the $n$-mode cat state, and $D_{n}=2^{(n-1)/2}$ is the upper bound of
the MK signal imposed by classical models. }
\end{figure}
For the 5-mode cat state, the MK signal based on the effective rotation
operators is given by%
\begin{eqnarray}
S_{5} &&=-E_{5}(0,-\pi /4,-\pi /4,-\pi /4,-\pi /4)+ \nonumber \\&&6E_{5}(0,-\pi /4,-\pi
/4,\pi /4,\pi /4)  -\nonumber \\&& E_{5}(0,\pi /4,\pi /4,\pi /4,\pi /4)-\nonumber \\&& 4E_{5}(\pi /2,\pi /4,\pi /4,\pi
/4,-\pi /4)  +\nonumber \\&& 4E_{5}(\pi /2,\pi /4,-\pi /4,-\pi /4,-\pi /4),
\end{eqnarray}%
where%
\begin{eqnarray}
E_{5}(\phi _{1},\phi _{2},\phi _{3},\phi _{4}) &=&{\cal N}_{5}^{2}\left\{
e^{-10\left\vert \alpha \right\vert ^{2}}+\prod_{j=1}^{5}K_{\alpha ,\alpha
}(\phi _{j})\right.  \nonumber \\
&&\left. +\left[ \prod_{j=1}^{5}K_{\alpha ,-\alpha }(\phi _{j})+c.c.\right]
\right\} .\nonumber \\&&
\end{eqnarray}%
The red dashed line in Fig. 3 denotes the rescaled 5-mode MK signal, defined
as $R_{5}=S_{5}/4$, versus $\alpha $. The results show $R_{5}$ reaches the
classical upper bound 1 and the lower bound $2\sqrt{2}$ for true 5-partite
entanglement at $\alpha =0.281$ and 0.64, respectively, and quickly approaches the
maximum of 4. For $\alpha =1$, $R_{5}\simeq 3.882$, close to the maximal value.
In contrast, with displaced parity operators, the same value of the MK
signal corresponds to $\alpha \simeq 3.927$.

\section{CONCLUSIONS}

In conclusion, we have analyzed quantum nonlocality for $n$ bosonic modes in
an entangled coherent state using rotated parity operators. Our results show
that the correlations obtained with this approach are close to those for the
n-qubit GHZ state, and the MK inequality can be nearly maximally violated
even when the average quantum number in each mode is only about 1. This is
in distinct contrast with the previous study based on displaced parity
operators, where the obtained MK signal for the 3-mode cat state even does
not exeed the lower bound for genuine 3-partite entanglement for this size.
With this framework, a nearly maximal inequality violation requires the
quantum number in each mode to be improved by about 15 times compared to the
present result. Since the decoherence rate of a cat state increases with the
size, our results are useful for experimentally investigating quantum
nonlocality for entanglement of quasiclassical states of multiple bosonic
modes.

This work was supported by the National Natural Science Foundation of China
under Grant No. 11874114 and No. 11674060.

\end{document}